\documentclass[oneocolumn,preprintnumbers,amsmath,amssymb]{revtex4}

\usepackage{amsmath,latexsym,amssymb,amsfonts}
\usepackage[pdftex]{color,graphicx}
\usepackage{bm}
\usepackage{gensymb}

\begin{document}


\title{\textbf{
Quantum black holes: inside and outside}}

\author{
Wei-Chen Lin$^{a}$\footnote{{\tt archennlin@gmail.com}},
Dong-han Yeom$^{b,c,d,e}$\footnote{{\tt innocent.yeom@gmail.com}}
and
Dejan Stojkovic$^{f}$\footnote{{\tt ds77@buffalo.edu}}
}

\affiliation{
$^{a}$Department of Science Education, Ewha Womans University, Seoul 03760, Republic of Korea\\
$^{b}$Department of Physics Education, Pusan National University, Busan 46241, Republic of Korea\\
$^{c}$Leung Center for Cosmology and Particle Astrophysics, National Taiwan University, Taipei 10617, Taiwan\\
$^{d}$Department of Physics and Astronomy, University of Waterloo, Waterloo, ON N2L 3G1, Canada\\ 
$^{e}$Perimeter Institute for Theoretical Physics,  Waterloo, ON N2L 2Y5, Canada\\
$^{f}$HEPCOS, Department of Physics, SUNY at Buffalo, Buffalo, NY 14260, USA
}

\begin{abstract}
For a unitary description of an evaporating black hole, one usually chooses the time slices that cover only outside of the event horizon, which is mostly problem-free because the event horizon is not encountered. However, is there any justification for avoiding time slices that cover inside the event horizon? To answer the question, we investigate the Wheeler-DeWitt equation, where the time slices can cover both inside and outside the event horizon. We find that one can reasonably construct a wave packet that covers outside, but the wave function must be annihilated near the event horizon. This observation strongly suggests that we cannot choose a coherent state for a spacelike hypersurface that crosses the event horizon. To explain the unitary time evolution, we must keep the slices as coherent states; hence, they must always be outside the event horizon. In contrast, inside the horizon, we cannot have a single coherent state of a classical spacetime. Hence, the interior must be a superposition of several coherent states, which implies that there exists a horizon-scale uncertainty and a black hole should be viewed as a highly quantum macroscopic object. We provide a synthetic approach to understanding the information loss paradox from this perspective.
\end{abstract}

\maketitle

\newpage

\tableofcontents

\section{Introduction}

Understanding the information loss paradox is one of the most critical issues in modern theoretical physics~\cite{Hawking:1976ra}. This reveals the tension between quantum mechanics and general relativity; hence, reconciling this tension will guide us to the fundamental understanding of the ultimate theory. To form a black hole, we need to consider gravitational collapse. After some matter carrying arbitrary amount of information collapses and forms a black hole, how can we take  out this information?

After discussing and debating black hole complementarity \cite{Susskind:1993if}, we can conclude that the following assumptions are inconsistent \cite{Yeom:2009zp,Almheiri:2012rt} (see also \cite{Hutchinson:2013kka}).
\begin{itemize}
\item[-- 1.] \textit{Unitarity} of the entire life of black hole evolution.
\item[-- 2.] \textit{Local quantum field theory} that explains Hawking radiation.
\item[-- 3.] \textit{General relativity} is satisfied all over the spacetime except for singularities.
\item[-- 4.] \textit{The Bekenstein-Hawking entropy is the Boltzmann entropy} of a black hole.
\item[-- 5.] \textit{There exists an observer} who can count information or entropy quantities.
\end{itemize}
Therefore, it is reasonable to conclude that if the unitarity is preserved and we are an observer who observes the unitary evolution of the black hole, then the Assumptions 2, 3, or 4 must be violated. In other words, a semi-classical description (local quantum field theory, general relativity, and known statistical expectations) is no longer valid in the bulk region. Therefore, we may need a non-perturbative description to understand the information loss paradox.

The most conservative way to understand this is to introduce the \textit{wave function of the universe} using the \textit{canonical approach} of quantum gravity \cite{DeWitt:1967yk}. This wave function is a functional of the metric and field variables on a 3-surface. The master equation that decides the wave function is known as the \textit{Wheeler-DeWitt equation} for gravity and fields.

If we know the Wheeler-DeWitt equation and the wave function on a 3-surface, how can we explain the conservation of information? The first thing that we can do is to introduce \textit{time} by introducing a clock of the unitary observer \cite{Page:1983uc}, and obtain the functional Schr{\"o}dinger equation \cite{Vachaspati:2006ki}. However, the choice of time is not sufficient. We also need a \textit{series of time slices}. If we choose 3-hypersurfaces only outside the horizon, then there is no effective event horizon. Hence, it is not too surprising that the time evolution is unitary and that all information will be recovered eventually \cite{Saini:2015dea,Baccetti:2016lsb,Ayon-Beato:1999kuh,Ashtekar:2005cj,Haggard:2014rza}. On the other hand, it has been argued that black holes might in fact be highly quantum macroscopic objects \cite{Dai:2020irc}. It is therefore reasonable to ask \textit{what will happen if we choose a hypersurface that crosses the event horizon}? In this paper, we primarily address the question: Can we select a 3-hypersurface that crosses the horizon? If it is not allowed, what is the physical reason?

Of course, due to the covariance of general relativity, in principle, we can choose any slice that crosses the event horizon; the hypersurface can even touch the singularity. In this paper, we will write down the Wheeler-DeWitt equation with the time slice that crosses over the event horizon. If the time slice presents a consistent classical boundary condition, e.g., an in-state at past infinity, then the wave function for the slice will correspond to a \textit{coherent state}. Now, the coherent state will evolve as time goes on, according to the functional Schr{\"o}dinger equation, which is manifestly unitary. However, we can ask whether its coherence will be preserved for all time. In this paper, we will demonstrate that coherence can break down if the time slice intersects the event horizon. This means that to explain the unitary evolution according to the asymptotic observer's clock, one needs to choose time slices only outside the event horizon.

Then, what is the meaning of such a \textit{decoherence} of the horizon scale? What will an infalling observer experience at the horizon? In this paper, we will address these questions within the orthodox framework of canonical quantum gravity.

The paper is organized as follows. In Sec.~\ref {sec:for}, we illustrate the formalism and fundamental questions in the context of the information loss paradox. The key idea is that we need to introduce a time and corresponding slicings that must be described by a coherent state. In Sec.~\ref{sec:qua}, we discuss the canonical quantization of a Schwarzschild black hole and derive the Wheeler-DeWitt equation. In Sec.~\ref{sec:ana}, we constructively analyze the Wheeler-DeWitt equation's solution. From this solution, we discuss their physical meaning in Sec.~\ref{sec:int}. Finally, in Sec.~\ref{sec:dis}, we summarize our findings and comment on future research topics.

\section{\label{sec:for}Formalism and fundamental problems}

In this section, we discuss the canonical quantum gravity as the formalism of our approach. To address the information loss paradox, we need to introduce time and time slices. The question is whether the time slice can be justified. To address this issue, we will emphasize the importance of a coherent state.

\subsection{Canonical quantum gravity}

Let us start with the most conservative approach to quantum gravity: the \textit{canonical approach} \cite{DeWitt:1967yk}. The master equation of this approach is the \textit{Wheeler-DeWitt equation}, which has the form:
\begin{eqnarray}
\hat{\mathcal{H}} \Psi [h_{ab}, \phi] = 0,
\end{eqnarray}
where $\hat{\mathcal{H}}$ is the quantum Hamiltonian constraint, $\Psi$ is the wave function of the Universe, which is a functional of the 3-hypersurface $h_{ab}$, and $\phi$ is a matter field on the hypersurface.

We assume that this master equation includes all the information, and hence, there must be no loss of information for evaporating black holes. However, it is fair to say that we need an evolution parameter, i.e. \textit{time}, to say anything about unitary \textit{time} evolution. On the other hand, the Wheeler-DeWitt equation in itself has apparently no time. Therefore, we need to introduce the concept of time before we ask the question about the unitary time evolution.

\subsection{Introducing time and the functional Schr{\"o}dinger equation}

There may be several approaches to introducing time (see e.g. \cite{Vachaspati:2006ki}), but when focusing on black holes, it is sufficient to follow the Page-Wootters formalism \cite{Page:1983uc}. We introduce a clock system which is supposed to be located at infinity; this divides the Hamiltonian as $\hat{\mathcal{H}} = \hat{\mathcal{H}}^{G}+\hat{\mathcal{H}}^{C}$, where $\hat{\mathcal{H}}^{G}$ is the Hamiltonian for the gravity and matter part, and $\hat{\mathcal{H}}^{C}$ for the clock part. By subtracting the degrees of freedom of the clock, we can further introduce a separation of variables:
\begin{eqnarray}
| \Psi \rangle = \int dt | t \rangle | \psi (t) \rangle,
\end{eqnarray}
where $| t \rangle$ acts on $\hat{\mathcal{H}}^{C}$ and $| \psi(t) \rangle$ acts on $\hat{\mathcal{H}}^{G}$. From this, we can introduce the time-dependent equation, the so-called functional Schr{\"o}dinger equation \cite{Vachaspati:2006ki}:
\begin{eqnarray}
\hat{\mathcal{H}}^{G} | \psi \rangle = i \frac{\partial}{\partial t} | \psi \rangle.
\end{eqnarray}

This is still not sufficient, because \textit{introducing time implies a series of time slices}. How can we choose time slices for a given clock; in other words, \textit{how can we choose a series of classical 3-hypersurfaces} corresponding to the time flow of the clock at infinity?

\begin{figure}[h]
\begin{center}
\includegraphics[scale=0.5]{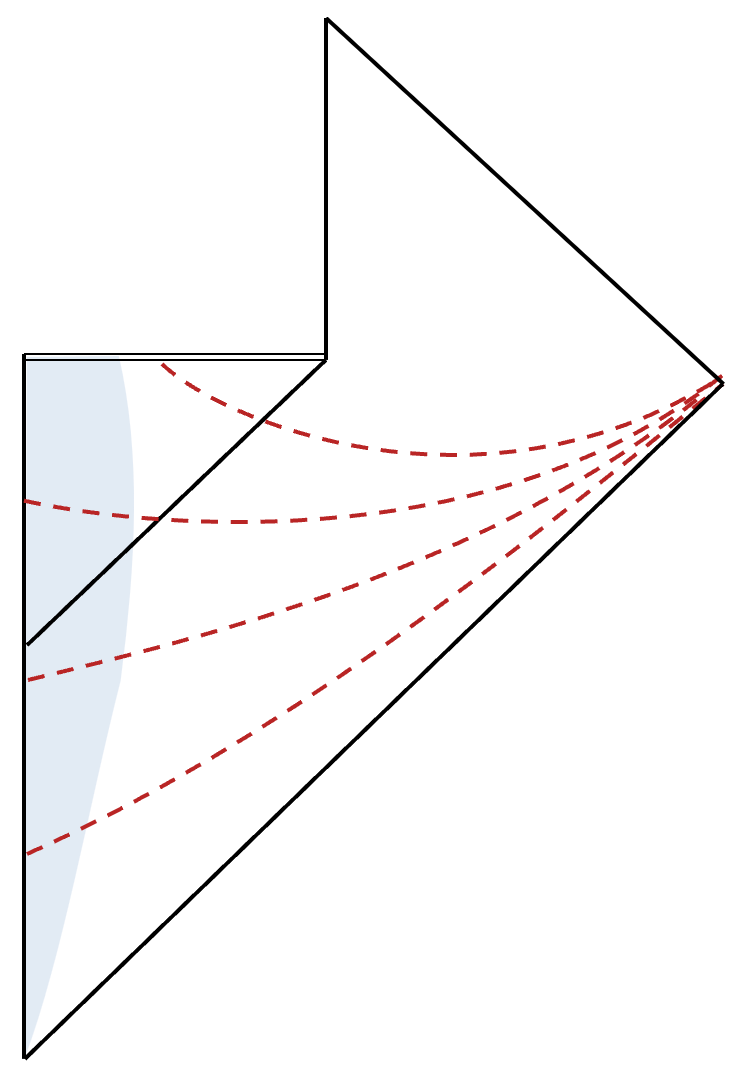}
\caption{\label{fig:typicalslicing}Typical time slices of a semi-classical black hole. Which slices are self-consistent in terms of canonical quantum gravity?}
\end{center}
\end{figure}

At first glance, it seems that we can choose any hypersurface due to general covariance (Fig.~\ref{fig:typicalslicing}). However, obviously, it is not true, because \textit{we do not know what the allowed hypersurface is unless we solve the Hamiltonian dynamics}. Therefore, there is an issue; to solve the Schr{\"o}dinger equation, we need to define time slices; but to define the time slices, we need to solve the Schr{\"o}dinger equation. So, how can we escape from this loop?

\subsection{Choice of time slices and the coherent state}

There is a way to escape the loop. Let us consider introducing time slices, assuming a classical spacetime. This is a reasonable guess, but as we have mentioned, unless we solve the Schr{\"o}dinger equation, we do not know whether we can do it or not. However, we can impose the \textit{consistency} condition, in the sense that such a classical choice of the hypersurface is still robust in the context of quantum gravity.

Then, what is the quantum consistency condition for a given 3-hypersurface? A typically accepted answer is that there exists a wave function for the 3-hypersurface as a \textit{coherent state}. If this coherent state is a solution of the Wheeler-DeWitt equation, the use of such a hypersurface is a robust way to describe the time evolution. On the other hand, if the coherence breaks down, then such a hypersurface choice is no longer consistent. Hence, this may indicate that a single classical hypersurface is not sufficient to carry the entire information. Therefore, when considering the unitary time evolution, we must choose a series of 3-hypersurfaces that coherent states can describe (as long as this is possible).

Now, how can we determine that a time slicing that crosses the event horizon is consistent with a coherent state? As we solve the Wheeler-DeWitt wave function, we will assign a Gaussian wave packet with a given dispersion. If the dispersion is preserved, then the classicality is preserved, and hence the wave packet describes a consistent coherent state. However, if the structure of the Gaussian wave packet is broken during the evolution, this is strong evidence that it is not possible to choose a coherent state along such a time slice. We will show the detailed computations in the following sections.

\section{\label{sec:qua}Quantization of a Schwarzschild black hole}

This section discusses the canonical quantization of a Schwarzschild black hole. There are examples of the Wheeler-DeWitt equation of a Schwarzschild black hole \cite{Kuchar:1994zk,Bojowald:2018xxu}, but in this paper, we will mainly focus on the time slicing that covers both inside and outside the event horizon.

\subsection{Model}

Let us consider Einstein gravity
\begin{eqnarray}
S = \int d^{4}x \sqrt{-g} \frac{\mathcal{R}}{16\pi},
\end{eqnarray}
where $\mathcal{R}$ is the Ricci scalar and $g_{\mu\nu}$ is the metric. We further assume the spherical symmetric metric ansatz
\begin{eqnarray}
ds^{2} = \alpha^{2}(t,x) \left( - dt^{2} + dx^{2} \right) + \frac{\beta^{2}(t,x)}{\alpha^{2}(t,x)} d\Omega_{2}^{2},
\end{eqnarray}
where $t$ and $x$ are coordinate variables, $\alpha$ and $\beta$ are functions of $t$ and $x$. This metric is useful for choosing regular metric parameters around the apparent horizon \cite{Waugh:1986jh,Nakonieczna:2018tih}. This is a generalization of the Kruskal-Szekeres coordinates.

By plugging this metric ansatz, integrating over the solid angle, and applying the integration by parts, we obtain
\begin{eqnarray}
S = \frac{1}{2} \int dt dx \left( \alpha^{2} + \frac{\beta^{2}}{\alpha^{4}} \left( \alpha_{,t}^{2} - \alpha_{,x}^{2} \right) - \frac{1}{\alpha^{2}} \left( \beta_{,t}^{2} - \beta_{,x}^{2} \right) \right).
\end{eqnarray}

\subsection{Classical Hamiltonian}

The Lagrangian is now
\begin{eqnarray}
L = \int dx \mathcal{L} = \frac{1}{2} \int dx \left( \alpha^{2} + \frac{\beta^{2}}{\alpha^{4}} \left( \alpha_{,t}^{2} - \alpha_{,x}^{2} \right) - \frac{1}{\alpha^{2}} \left( \beta_{,t}^{2} - \beta_{,x}^{2} \right) \right).
\end{eqnarray}
One can define the canonical momenta:
\begin{eqnarray}
p_{\alpha} &\equiv& \frac{\partial \mathcal{L}}{\partial \alpha_{,t}} = \frac{\beta^{2} \alpha_{,t}}{\alpha^{4}},\\
p_{\beta} &\equiv& \frac{\partial \mathcal{L}}{\partial \beta_{,t}} = - \frac{\beta_{,t}}{\alpha^{2}}.
\end{eqnarray}
From these relations, one can find that
\begin{eqnarray}
\alpha_{,t} &=& \frac{\alpha^{4}}{\beta^{2}} p_{\alpha},\label{eq:conv1}\\
\beta_{,t} &=& - \alpha^{2} p_{\beta}.\label{eq:conv2}
\end{eqnarray}

Now, the Hamiltonian $H$ is obtained after the Legendre transformation:
\begin{eqnarray}
H &\equiv& \int dx \left( p_{\alpha} \alpha_{,t} + p_{\beta} \beta_{,t} \right) - L\\
&=& - \frac{1}{2} \int dx  \left( \alpha^{2} - \frac{\beta^{2}}{\alpha^{4}} \left( \alpha_{,t}^{2} + \alpha_{,x}^{2} \right) + \frac{1}{\alpha^{2}} \left( \beta_{,t}^{2} + \beta_{,x}^{2} \right)\right) \\
&=& \int dx  \left( \frac{\alpha^{4}}{2\beta^{2}} p_{\alpha}^{2} - \frac{\alpha^{2}}{2} p_{\beta}^{2} - \frac{1}{2} \left( \alpha^{2} - \frac{\beta^{2}}{\alpha^{4}} \alpha_{,x}^{2} +\frac{\beta_{,x}^{2}}{\alpha^{2}} \right) \right)\\
&\equiv& \int dx \mathcal{H},
\end{eqnarray}
where $\mathcal{H}$ is the Hamiltonian density.

\subsection{Wheeler-DeWitt equation}

To obtain the quantum Hamiltonian constraint, we introduce the commutation relations:
\begin{eqnarray}
\left[ \alpha, p_{\alpha} \right] = \left[ \beta, p_{\beta} \right] = i,
\end{eqnarray}
and hence,
\begin{eqnarray}
p_{\alpha} &\rightarrow& \frac{\delta}{i \delta \alpha}, \\
p_{\beta} &\rightarrow& \frac{\delta}{i \delta \beta}.
\end{eqnarray}

One may define $a \equiv \log \alpha$ and $b \equiv \log \beta$. By choosing a reasonable operator ordering (such as \cite{Bouhmadi-Lopez:2019kkt,Kang:2022tkb}), we obtain the Wheeler-DeWitt equation:
\begin{eqnarray}
\left( - \frac{\delta^{2}}{\delta a^{2}} + \frac{\delta^{2}}{\delta b^{2}} - \left( e^{2b} - e^{4(b-a)} \left( a_{,x}^{2} - b_{,x}^{2} \right)\right) \right) \psi[a,b] = 0.
\end{eqnarray}
In the potential term, $x$-dependence exists. Therefore, a differential equation is not well-defined; rather, it is a \textit{functional} differential equation. This behavior is quite typical in the spherically symmetric examples \cite{Kuchar:1994zk,Bojowald:2018xxu,Demers:1995tr}. Here, the potential of the Hamiltonian is
\begin{eqnarray}
\mathcal{V}\left[a,b ; x \right] = e^{2b} - e^{4(b-a)} \left( a_{,x}^{2} - b_{,x}^{2} \right).
\end{eqnarray}


\begin{figure}
\begin{center}
\includegraphics[scale=0.7]{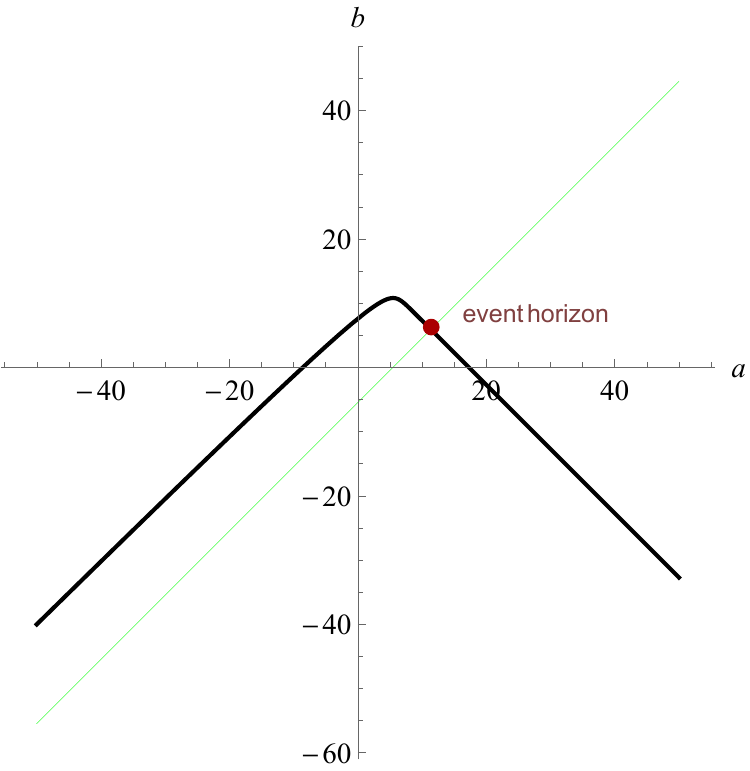}
\caption{\label{fig:cansp}Classical trajectory of $a$ and $b$ space (black) for $M = 100$. The green-colored line denotes $r = 2M$. The left end corresponds to infinity, while the right end corresponds to a singularity.}
\end{center}
\end{figure}

\section{\label{sec:ana}Analysis of the wave functions}

Now, let us analyze the solution of the Wheeler-DeWitt equation. It is impossible to solve the equation directly, but there is an intuitive way to understand the nature of the solution. In addition, we are opening up the possibility of a new interpretation.

\subsection{Classical analysis}

The Kruskal-Szekeres coordinate is
\begin{eqnarray}
\alpha^{2} = \frac{32M^{3}}{r} e^{-r/2M}
\end{eqnarray}
and
\begin{eqnarray}
\frac{\beta}{\alpha} = r = 2M \left( 1 + \mathcal{W}_{0} \left( \frac{x^{2} - t^{2}}{e} \right) \right),
\end{eqnarray}
where $\mathcal{W}_{0}$ is the product logarithm ($\mathcal{W}_{0}(z)$ is the principal branch of solutions of the equation $\mathcal{W}e^{\mathcal{W}} = z$). Therefore,
\begin{eqnarray}\label{eq:clconst}
a = \log 32M^{3} - b - \frac{e^{b-a}}{2M}
\end{eqnarray}
is the relation of the on-shell condition in the phase space. In addition, we obtain
\begin{eqnarray}
a(x,t) &=& \frac{1}{2} \left( \log 32M^{3} - \log r(x,t) - \frac{r(t,x)}{2M} \right),\\
b(x,t) &=& \frac{1}{2} \left( \log 32M^{3} + \log r(x,t) - \frac{r(x,t)}{2M} \right).
\end{eqnarray}
From this, one can easily check that $a \rightarrow  \infty$ and $b \rightarrow - \infty$ corresponds to the singularity ($r = 0$), while $a, b \rightarrow - \infty$ corresponds to $r = \infty$.

\subsection{Free propagation}

To exhibit (close to) classical behavior, the wave function has to peak along one specific classical trajectory (Eq.~(\ref{eq:clconst})) in the configuration space specified by the mass $M$ of a Schwarzschild solution. Due to the hyperbolic feature of the Wheeler-DeWitt equation, we can draw an analogy to the scattering of propagating waves in $1+1$ dimensions and provide a qualitative analysis of the existence of a classicalized wave function of the Wheeler-DeWitt equation as follows. 

Since the Hamiltonian's potential is negligible far away or deep inside the event horizon, we must consider the wave packets in these regions as free fields.
That is, if $|\mathcal{V}| \ll 1$, the Wheeler-DeWitt equation is approximately given by
\begin{eqnarray}
\left( - \frac{\delta^{2}}{\delta a^{2}} + \frac{\delta^{2}}{\delta b^{2}} \right) \psi[a,b] \simeq 0.
\end{eqnarray}

Now, note that the classical trajectory is given by Eq.~(\ref{eq:clconst}). According to this relation, in the $a \gg 0$ limit, the approximation $a+b \simeq \log 32M^{3}$ is satisfied. In the limit $a \ll 0$, the trajectory is approximately $a + e^{b-a}/2M \simeq \log 32M^{3}$. Note that $b - a$ must vary slowly, and hence in the limit $a \ll 0$, the classical trajectory is the $b - a \simeq \mathrm{const.}$ line. Based on this, we can say that away from the event horizon, the wave function showing strong classical behavior must be described by the right-moving wave $f(b-a)$ peaking at $b-a \simeq \mathrm{const.}$ in the scattering analogy. Meanwhile, when deep inside the event horizon, the wave function describing a classical Schwarzschild spacetime must correspond to the right-moving wave $g(b+a)$ with a peak at $a+b \simeq \log 32M^{3}$. Notice that both of them are right-moving due to the relation of $b$ and $r$ mentioned at the end of the previous subsection.  

Therefore, if $|\mathcal{V}| \ll 1$, the solutions of the free propagating modes are superpositions of
\begin{eqnarray}
\psi = e^{\pm ik(a-b)}
\end{eqnarray}
or
\begin{eqnarray}
\psi = e^{\pm ik(a+b)}.
\end{eqnarray}
The superposition of the former functions is for outside the event horizon, while that of the latter functions is for inside the event horizon. We need to construct a coherent state for $\mathcal{V} \ll 1$ region to describe a wave function for a classical hypersurface.

\subsection{Approximated potential}

If we consider a wave packet that has the peak at the classical solution, one can approximate the effective potential for a given time $t_{0}$ as follows:
\begin{eqnarray}
\mathcal{V}\left[a,b ; x \right]  = e^{2b} - e^{4 (b-a)} \left( \frac{4x^{2}\mathcal{W}_{0}\left((x^{2}-t_{0}^{2})/e\right)^{2} }{\left(x^{2} - t_{0}^{2}\right)^{2} \left(1 + \mathcal{W}_{0}\left((x^{2}-t_{0}^{2})/e\right) \right)^{3}} \right).
\end{eqnarray}
The question is how we can present this as a function of $a$ and $b$. There might be no unique way, but we use the prescription:
\begin{eqnarray}
\mathcal{W}_{0}\left((x^{2}-t_{0}^{2})/e\right) &\rightarrow& \frac{e^{b-a}}{2M} - 1,\\
x^{2} - t_{0}^{2} &\rightarrow& - 32 M^{3} \left( 1 - \frac{e^{b-a}}{2M} \right) e^{-(a+b)}.
\end{eqnarray}
We finally obtain
\begin{eqnarray}\label{eq:standard}
\mathcal{V}_{\mathrm{eff}}\left[a,b \right]  = 2 e^{2b} - \frac{t_{0}^{2}}{32 M^{3}} e^{a + 3b} - \frac{1}{2M} e^{3b-a}.
\end{eqnarray}

It is essential to specify the validity regime of this effective presentation. First, if $|\mathcal{V}| \ll 1$, then this potential is a good approximation. In addition, along the classical contour, this is also a good approximation. What we will consider is the free propagation of the wave function in $|\mathcal{V}| \ll 1$ regime, and its collision with the potential barrier (or well). Therefore, if we observe right before the $|\mathcal{V}| \gg 1$ region, our effective description is still valid.

\subsection{Structure of the approximated potential}

\begin{figure}
\begin{center}
\includegraphics[scale=0.4]{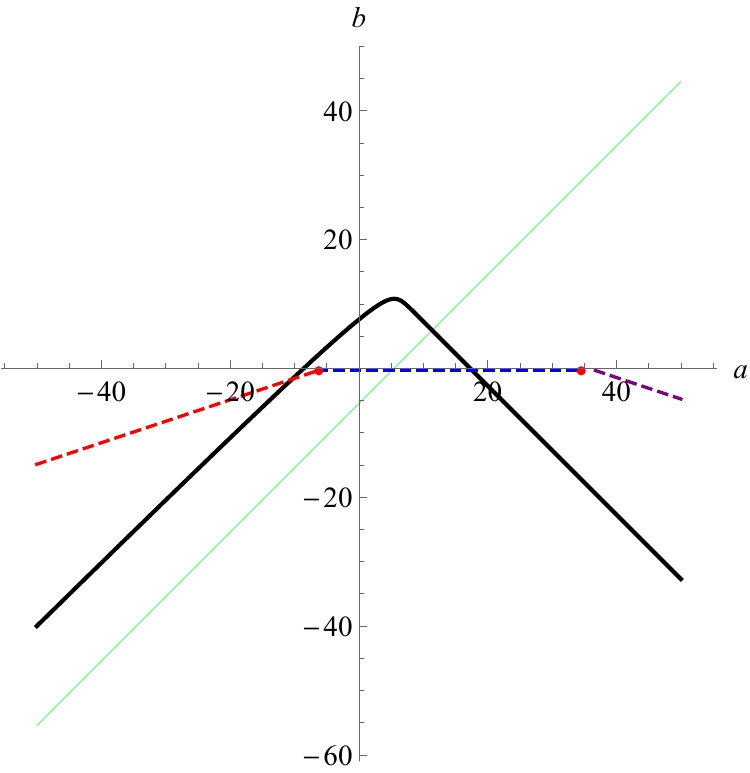}
\includegraphics[scale=0.4]{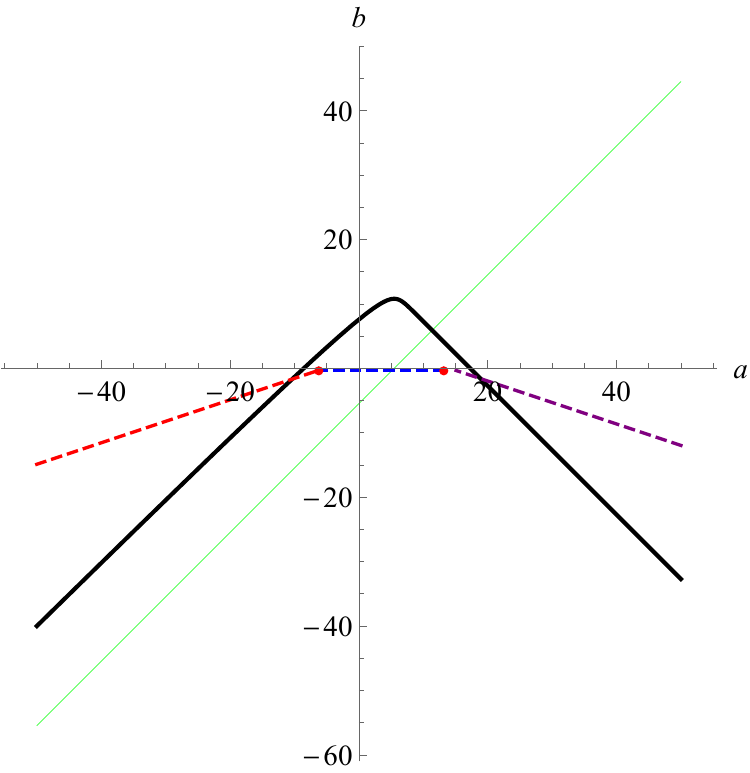}
\includegraphics[scale=0.4]{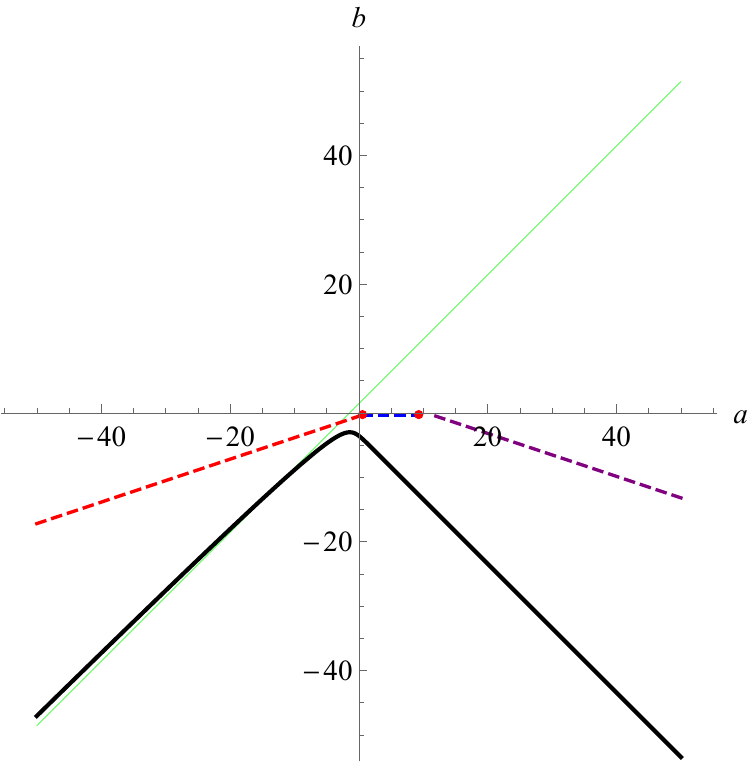}
\caption{\label{fig:earlytime}Several cases of potential boundaries. Left: the early time limit ($M = 100$, $t_{0} = 0.0001$). Middle: the late time limit ($M = 100$, $t_{0} = 5$). Right: the small mass limit ($M = 0.1$, $t = 0.001$). The red and violet dashed curves are potential wells, while the blue dashed curve is the potential barrier.}
\end{center}
\end{figure}

One can rewrite the effective potential as follows:
\begin{eqnarray}
\mathcal{V}_{\mathrm{eff}}\left[a,b \right]  = e^{2b + \log 2} - e^{a +3b + \log \frac{t_{0}^{2}}{32M^{3}}} - e^{3b-a+\log \frac{1}{2M}}.
\end{eqnarray}
Therefore, there are three exponentials, where the first term is the potential barrier, while the second and third terms are potential wells. The boundaries of each exponential are
\begin{eqnarray}
2b + \log 2 &=& 0,\\
a + 3b + \log \frac{t_{0}^{2}}{32M^{3}} &=& 0,\\
3b - a + \log \frac{1}{2M} &=& 0,
\end{eqnarray}
respectively. In Fig.~\ref{fig:earlytime}, we denote the boundaries using a blue dashed line, a purple dashed line, and a red dashed line, respectively.

The crossing point between the red and the blue dashed line is at
\begin{eqnarray}
(a_{1}, b_{1}) = \left( \log \left( \frac{1}{4 \sqrt{2}M} \right), - \frac{1}{2} \log 2 \right),
\end{eqnarray}
while the crossing point between the blue and the purple dashed line is at
\begin{eqnarray}
(a_{2}, b_{2}) = \left( \log \left( \frac{8\sqrt{2} M^{3}}{t_{0}^{2}} \right), - \frac{1}{2} \log 2 \right).
\end{eqnarray}
These two points are denoted by red dots in Fig.~\ref{fig:earlytime}.

From this analysis, we can notice several points as follows:
\begin{itemize}
\item[--] The classical contour from outside the horizon touches the red dashed line (left and middle of Fig.~\ref{fig:earlytime}); hence, the wave packet for outside the event horizon touches the exponential potential well. This does not depend on the choice of time.
\item[--] The classical contour from inside the horizon touches either the blue dashed line or the purple dashed line (left and middle of Fig.~\ref{fig:earlytime}, respectively). In the early time limit, the wave packet inside the event horizon encounters the exponential potential barrier (left of Fig.~\ref{fig:earlytime}); in the late time limit, the wave packet inside the event horizon interacts with the exponential potential well (middle of Fig.~\ref{fig:earlytime}).
\item[--] If we consider the sub-Planckian mass black hole, the classical contour does not touch any potential barrier or well (right of Fig.~\ref{fig:earlytime}). 
\end{itemize}

\subsection{Wave function of the potential well and barrier}

Typically, the wave packets from the region $|\mathcal{V}| \ll 1$ (infinity or singularity) will reach either the potential barrier or the potential well. Let us analyze their response to them.

\subsubsection{Analytic examples of the exponential potential well or barrier}

Before we deal with the approximated potential, let us demonstrate an analytic example of the exponential potential well or barrier. One can present the model as follows:
\begin{eqnarray}
\left( - \frac{\partial^{2}}{\partial a^{2}} + \frac{\partial^{2}}{\partial b^{2}} - \Lambda e^{\beta b} \right) \psi[a,b] = 0,
\end{eqnarray}
where $\beta > 0$ is a constant and $\Lambda$ is a constant. If $\Lambda > 0$, this becomes a potential barrier (see also \cite{Bouhmadi-Lopez:2019kkt}), while if $\Lambda < 0$, this becomes a potential well (see also \cite{Chien:2023kqw}).

In this case, one can introduce the separation of variables: $\psi[a,b] = A(a) B(b)$, where
\begin{eqnarray}
A'' + k^{2} A &=& 0,\\
B''+ \left( k^{2} - \Lambda e^{\beta b} \right) B &=& 0.
\end{eqnarray}
The generic solution of $A(a)$ is
\begin{eqnarray}
A(a) = e^{\pm ik a}.
\end{eqnarray}
The generic solution of $B(b)$ can be classified as follows:
\begin{itemize}
\item[--] If $\Lambda > 0$ (potential barrier), the generic solution is
\begin{eqnarray}
B(b) = C_{1}  K_{2ik/\beta} \left( \frac{2 \sqrt{\Lambda} e^{\beta b/2}}{\beta} \right) + C_{2} I_{2ik/\beta} \left( \frac{2 \sqrt{\Lambda} e^{\beta b/2}}{\beta} \right).
\end{eqnarray}
Note that the modified Bessel function $I_{2ik/\beta}$ diverges in the large $b$ limit, and hence, to assure the bounded wave function, we choose $C_{2} = 0$.
\item[--] If $\Lambda < 0$ (potential well), the generic solution is
\begin{eqnarray}
B(b) = C_{1}  J_{2ik/\beta} \left( \frac{2 \sqrt{|\Lambda|} e^{\beta b/2}}{\beta} \right) + C_{2} Y_{2ik/\beta} \left( \frac{2 \sqrt{|\Lambda|} e^{\beta b/2}}{\beta} \right).
\end{eqnarray}
Again, the Bessel functions of the second kind $Y_{2ik}$ diverge in the small $b$ limit, and hence, to assure a bounded wave function, we choose $C_{2} = 0$.
\end{itemize}

From this, one can construct a Gaussian wave packet for the $b \rightarrow - \infty$ limit. Again, we can separate the two cases.
\begin{itemize}
\item[--] If $\Lambda > 0$ (potential barrier), one can construct the Gaussian wave packet:
\begin{eqnarray}
\psi[a, b] &=& \int_{-\infty}^{\infty} \frac{2A e^{- \frac{\sigma^{2}k^{2}}{2}}}{\Gamma\left[-\frac{2ik}{\beta}\right]} \left( \frac{\sqrt{\Lambda}}{\beta} \right)^{-i\frac{2k}{\beta}} e^{-ika} K_{2ik/\beta} \left( \frac{2 \sqrt{\Lambda} e^{\beta b/2}}{\beta} \right) dk \\
&\underset{b\rightarrow -\infty}\simeq& \int_{-\infty}^{\infty} A e^{- \frac{\sigma^{2}k^{2}}{2}} \left( e^{-ik (a-b)} + \frac{\Gamma\left[\frac{2ik}{\beta}\right] \left( \frac{\sqrt{\Lambda}}{\beta} \right)^{-i\frac{2k}{\beta}}}{\Gamma\left[-\frac{2ik}{\beta}\right] \left( \frac{\sqrt{\Lambda}}{\beta} \right)^{i\frac{2k}{\beta}}} e^{-ik (a+b)} \right) dk.
\end{eqnarray}
Therefore, for the $b \rightarrow - \infty$ limit, there must be two wave packets, where one is approaching the barrier and the other is its mirror image bounced by the potential barrier.
\item[--] If $\Lambda < 0$ (potential well), one can construct the Gaussian wave packet:
\begin{eqnarray}
\psi[a, b] &=& \int_{-\infty}^{\infty} A e^{- \frac{\sigma^{2}k^{2}}{2}} \Gamma\left[ 1+\frac{2ik}{\beta} \right] {\left( \frac{\sqrt{|\Lambda|}}{\beta} \right)^{-i\frac{2k}{\beta}}} e^{-ika} J_{2ik/\beta} \left( \frac{2 \sqrt{|\Lambda|} e^{\beta b/2}}{\beta} \right) dk \\
&\underset{b\rightarrow -\infty}\simeq& \int_{-\infty}^{\infty} A e^{- \frac{\sigma^{2}k^{2}}{2}} e^{-ik (a-b)} dk.
\end{eqnarray}
Therefore, in the potential well case, the wave packet can be entirely absorbed without bouncing back. 
\end{itemize}
In general, we can physically conclude that if a wave packet reaches the exponential potential barrier, we can expect a perfect bounce; on the other hand, if a wave packet reaches the exponential potential well, we can expect that the wave packet will be perfectly absorbed. Even though the potential structure is somewhat modified in realistic situations, we expect that this physical intuition is correct.

\subsubsection{Wave packets for outside}

Approximately, one can solve the equation:
\begin{eqnarray}
\left( - \frac{\partial^{2}}{\partial a^{2}} + \frac{\partial^{2}}{\partial b^{2}} + \frac{1}{2M} e^{3b - a} \right) \psi[a,b] = 0.
\end{eqnarray}

\begin{figure}[h]
\begin{center}
\includegraphics[scale=0.7]{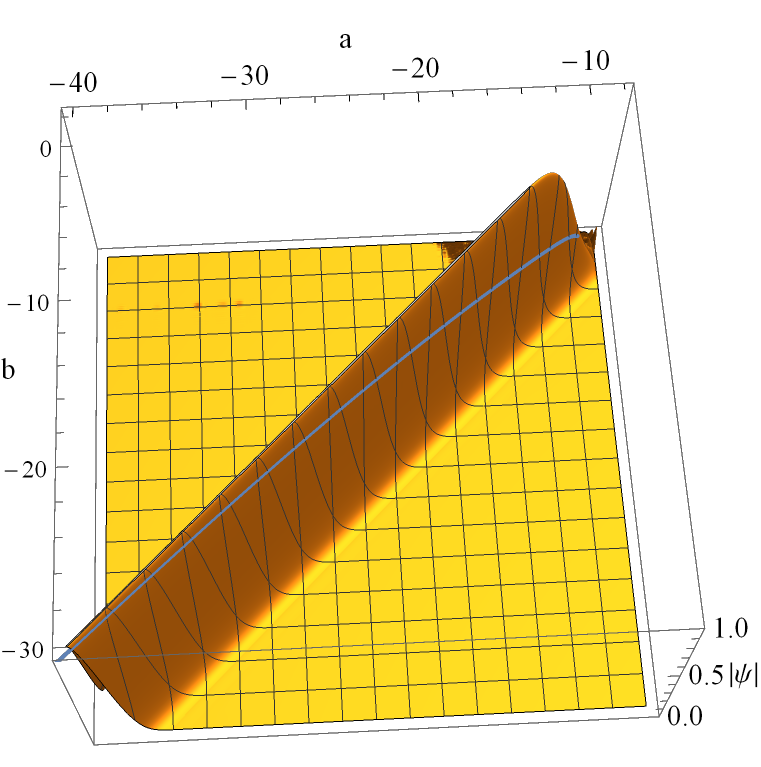}
\caption{\label{fig:sol1} A numerical solution for a wave packet starting from outside the event horizon. Here, we give $a_{m} = -100$, $a_{0} = -50$, $A = 1$, $\sigma = 1$, $M = 100$, and $a_{M} = -8.1$. One can provide a vanishing boundary condition at the potential well. The blue contour is the classical trajectory.}
\end{center}
\end{figure}

One can provide the boundary condition as follows, where our integration domain is $(-a_{m}, a_{M}) \times (-b_{m}, b_{M})$:
\begin{itemize}
\item[--] We introduce a Gaussian wave packet at $\psi[a, -b_{m}]$:
\begin{eqnarray}
\psi[a, -b_{m}] = A e^{- \frac{(a - a_{0})^{2}}{2 \sigma^{2}}},
\end{eqnarray}
where $A$ is a normalization constant, $\sigma$ is the standard deviation, and we choose $b_{m} = b_{0}$, where $(a_{0}, b_{0})$ corresponds a point on the classical trajectory. Here, we assume $- a_{m} \ll a_{0}$ and $\sigma$ is sufficiently small to prepare a localized wave packet.
\item[--] To provide a wave following the contour $a - b = \mathrm{const.}$, we provide the condition
\begin{eqnarray}
\left.\frac{\partial}{\partial b}\psi[a, b] \right| _{b = -b_{m}} = - \left.\frac{\partial}{\partial a} \psi[a, b]\right| _{b = -b_{m}}.
\end{eqnarray}
This is a good approximation for $a_{0}$, $b_{0} \rightarrow - \infty$.
\item[--] For the boundaries at $\psi[a_{m},b]$ and $\psi[a_{M},b]$, we provide
\begin{eqnarray}
\psi[a_{m},b] &=& \psi[a_{m},b_{m}],\\
\psi[a_{M},b] &=& \psi[a_{M},b_{m}].
\end{eqnarray}
If $\sigma$ is small enough, then $\psi[a_{m},b] \simeq \psi[a_{M},b] \simeq 0$. By adjusting $a_{M}$, one can provide the smoothly vanishing boundary condition for the large $b$ limit.
\end{itemize}

A numerical result is in Fig.~\ref{fig:sol1}. This is consistent with the perfect absorption case of the previous subsection. One can provide a boundary condition such that the wave packet smoothly vanishes as it approaches the potential well. This analysis relies on the approximation form of the potential, which is still a good approximation because the wave function mainly moves in the $|\mathcal{V}| \ll 1$ domain. After providing the boundary condition, the wave packet never passes over the potential well, where the validity may break down. Our wave packet lives within the valid regime of the approximated potential.

\subsubsection{Wave packets for inside: early time}

In the early time limit, i.e., if $t_{0} \lesssim 1$, the classical trajectory touches the potential barrier, and the equation is approximately as follows:
\begin{eqnarray}
\left( - \frac{\partial^{2}}{\partial a^{2}} + \frac{\partial^{2}}{\partial b^{2}} - 2 e^{2b} \right) \psi[a,b] = 0.
\end{eqnarray}
We know that this is precisely the case that we can solve analytically (with $\Lambda = 2$ and $\beta = 2$).

\begin{figure}[h]
\begin{center}
\includegraphics[scale=0.5]{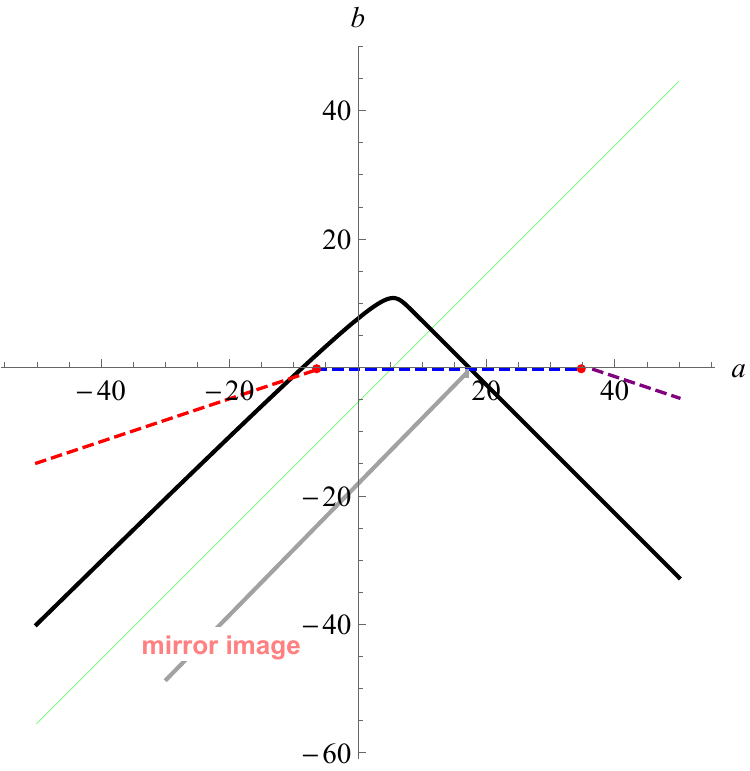}
\includegraphics[scale=0.45]{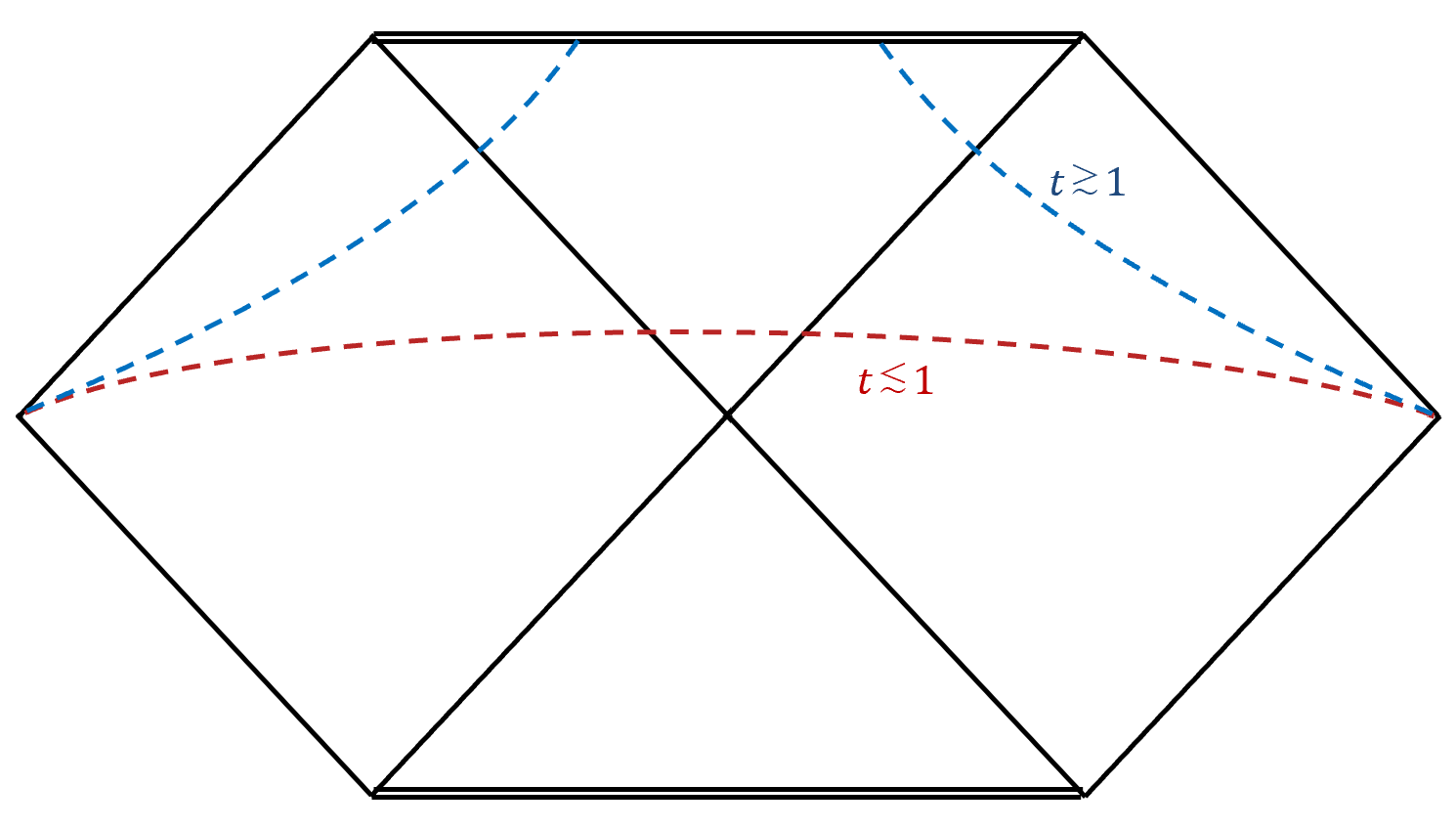}
\caption{\label{fig:refl}Left: If we consider a wave packet inside the horizon for early time, there must exist its mirror image that may correspond to a sub-Planckian mass limit geometry. Right: The early Kruskal slicing does not touch the singularity. Hence, it is not surprising that there is no consistent classical wave packet near the singularity for early time slicings.}
\end{center}
\end{figure}

Therefore, for a given classical path inside the horizon, $a + b = \mathrm{const.}$, that approaches the potential barrier, there must be a mirror image along the $a - b = \mathrm{const}.$ contour (left of Fig.~\ref{fig:refl}). Comparing with the right of Fig.~\ref {fig:earlytime}, this contour along the mirror image corresponds to the outside geometry of a sub-Planckian mass black hole. Therefore, this contour cannot be an on-shell trajectory, and we can conclude that in the early time, there is no classical interior. This conclusion is not surprising, as the Kruskal slicing of early times does not intersect the singularity (right of Fig.~\ref{fig:refl}).

\subsubsection{Wave packets for inside: late time}

Finally, in the late time, a wave packet can exist that covers the inside. The classical trajectory touches the potential well, and the equation is approximately as follows:
\begin{eqnarray}
\left( - \frac{\partial^{2}}{\partial a^{2}} + \frac{\partial^{2}}{\partial b^{2}} +\frac{t_{0}^{2}}{32 M^{3}} e^{a + 3b} \right) \psi[a,b] = 0.
\end{eqnarray}
Since this is a potential well again, we can expect a wave packet to be perfectly absorbed at the potential well; in other words, one can provide a vanishing boundary condition at the potential well.

\subsection{Domain where coherent states are allowed}

Finally, we can conclude that there is a domain where we can assign a coherent state. If the Gaussian wave packet maintains its shape, we can consider this region to correspond to a coherent state. Conversely, if we cannot continuously assign a Gaussian wave packet that is connected from infinity, we cannot consistently describe this region as a coherent state.

\begin{figure}[h]
\begin{center}
\includegraphics[scale=0.5]{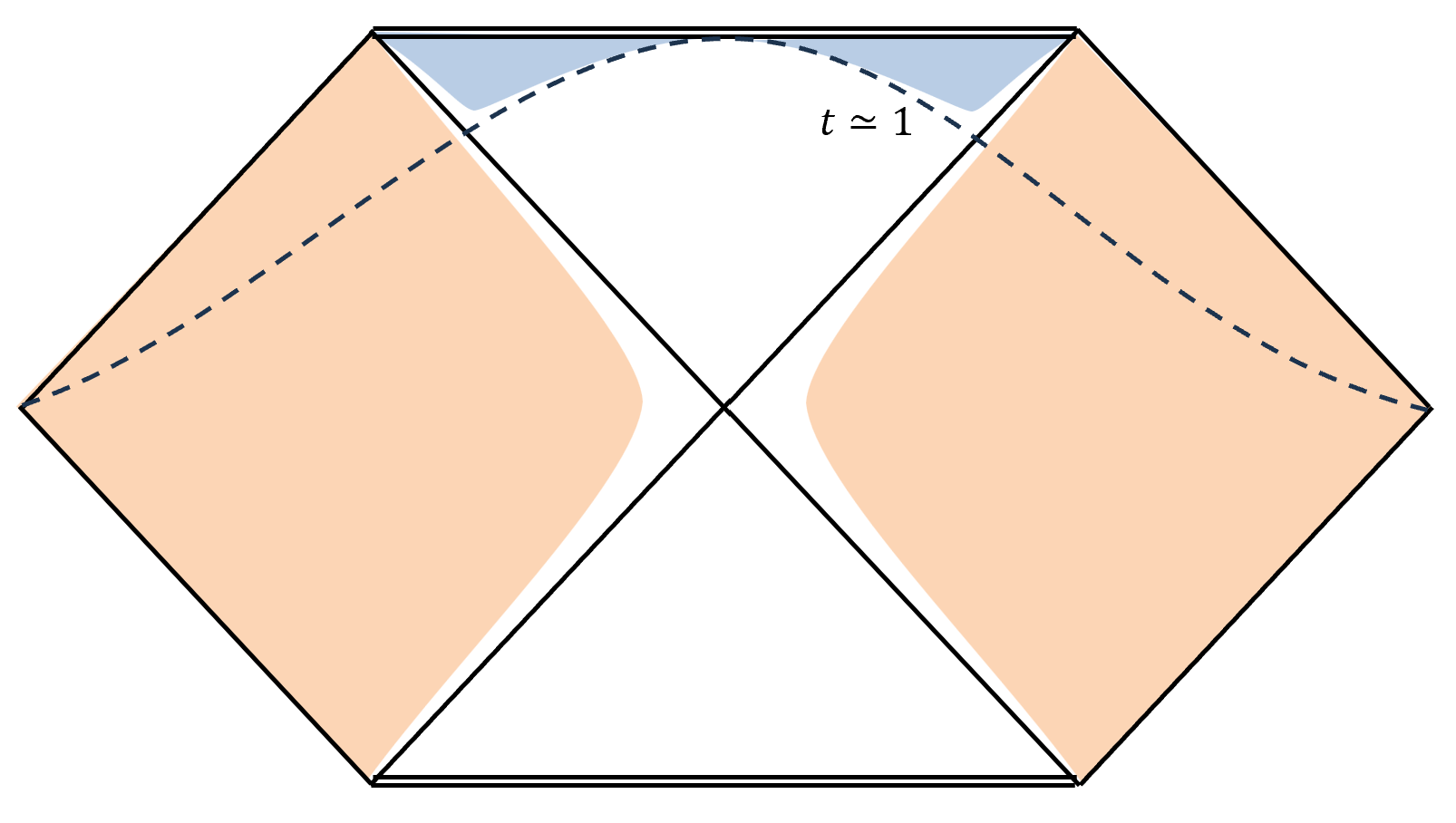}
\caption{\label{fig:coherentdom}Domain of the Schwarzschild spacetime where we can assign a coherent wave packet. The orange-colored region corresponds to outside the event horizon, and the blue-colored region corresponds to inside the event horizon. There is no slicing as a coherent state that smoothly connects from outside to inside the black hole.}
\end{center}
\end{figure}

Fig.~\ref{fig:coherentdom} presents our conclusion. One can assign a Gaussian wave packet, or a coherent state, outside the event horizon (orange-colored region), and it will be terminated slightly outside the horizon, because the wave packet smoothly vanishes near the potential well outside the event horizon. Since we observed that late-time slices can allow a Gaussian wave packet to exist inside the horizon, some parts near the singularity can be described as a coherent state (blue-colored region). However, two regions are disconnected. Therefore, if one starts the time slice from outside the horizon, e.g., at past infinity, then \textit{there is no slicing as a coherent state that smoothly connects from outside to inside the black hole}.

\section{\label{sec:int}Interpretations}

From the solutions of the Wheeler-DeWitt equation, we could conclude that the coherent state condition can only be assigned for the time slice outside the horizon. For a given time slice, the wave function seems to vanish at a surface outside the event horizon. Therefore, practically, there is a \textit{horizon scale black hole avoidance} in terms of the Wheeler-DeWitt wave function. The classical spacetime is annihilated to nothing at the horizon scale \cite{Bouhmadi-Lopez:2019kkt}, but how can we interpret this? What happens if an observer falls into the horizon? We should just follow what the wave function is telling us.

\subsection{Physical meaning of the \textit{annihilation-to-nothing}}

The wave packet is annihilated at the horizon due to the perfect absorption near the event horizon. Then, should we interpret this as if there was no physical interior of the black hole? To answer this question correctly, it will be better to see another physical example that has a similar physical behavior.

\subsubsection{De Sitter example}

Let us consider the action
\begin{eqnarray}
S = \int \sqrt{-g} dx^{4} \left[ \frac{1}{16\pi} \left( R - 2 \Lambda \right) - \frac{1}{2} \left( \partial \phi \right)^{2} \right],
\end{eqnarray}
where $\phi$ is a free scalar field. In this context, let us think about a de Sitter space with the flat slicing:
\begin{eqnarray}
ds^{2} = - N^{2} (t) dt^{2} + a^{2}(t) \left( dx^{2} + dy^{2} + dz^{2} \right).
\end{eqnarray}
The reduced action is
\begin{eqnarray}
S = \mathcal{V} \int dt N \left[ - \frac{3}{8\pi} \frac{a \dot{a}^{2}}{N^{2}} - a^{3} \frac{\Lambda}{8\pi} + \frac{1}{2} a^{3} \frac{\dot{\phi}^{2}}{N^{2}} \right],
\end{eqnarray}
where $\mathcal{V}$ is the spatial volume. The generalized momenta of $a$ and $\phi$ are
\begin{eqnarray}
p_{a} &=& - \frac{3 a \dot{a}}{4\pi N},\\
p_{\phi} &=& \frac{a^{3} \dot{\phi}}{N}.
\end{eqnarray}
Therefore, the Hamiltonian density is
\begin{eqnarray}
\mathcal{H} = \dot{a} p_{a} + \dot{\phi} p_{\phi} - \mathcal{L} = \frac{N}{8\pi} \left( - \frac{16 \pi^{2}}{3} \frac{p_{a}^{2}}{a} + \frac{4\pi p_{\phi}^{2}}{a^{3}}+ \Lambda a^{3} \right).
\end{eqnarray}

From this, the Wheeler-DeWitt equation is (up to the operator ordering)
\begin{eqnarray}
\hat{\mathcal{H}} \psi[\alpha, \varphi] = \left( \frac{\partial^{2}}{\partial \alpha^{2}} - \frac{\partial^{2}}{\partial \varphi^{2}} + \Lambda e^{24\pi \alpha/\sqrt{3}} \right) \psi [\alpha, \varphi] = 0,
\end{eqnarray}
where
\begin{eqnarray}
\alpha &\equiv& \frac{\sqrt{3}}{4\pi} \log a, \\
\varphi &\equiv& \frac{1}{\sqrt{4\pi}} \phi.
\end{eqnarray}
This is exactly the same as the equation with the exponential potential well that we have discussed. The solution with a Gaussian wave packet can be presented as follows (here, $\Lambda \equiv 3 H^{2}$):
\begin{eqnarray}
\psi[\alpha, \varphi] = \int_{-\infty}^{\infty} A e^{- \frac{\sigma^{2}k^{2}}{2}} \left( \frac{H}{8\pi} \right)^{\frac{ik}{4\pi\sqrt{3}}} \Gamma\left[ 1 - \frac{ik}{4\pi\sqrt{3} } \right] J_{-\frac{ik}{4\sqrt{3}\pi}} \left( \frac{H e^{4\sqrt{3}\pi \alpha}}{4\pi} \right) e^{i k \varphi} dk.
\end{eqnarray}

\begin{figure}[h]
\begin{center}
\includegraphics[scale=0.6]{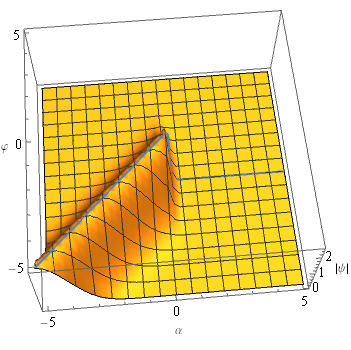}
\includegraphics[scale=0.7]{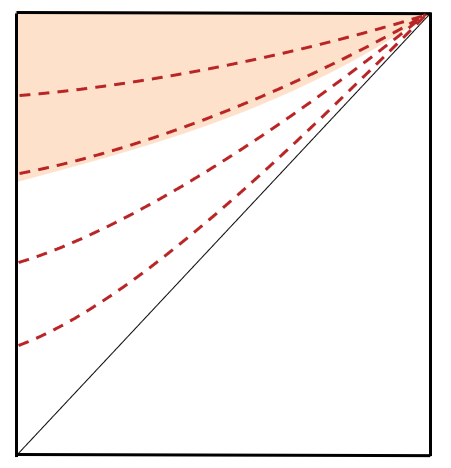}
\caption{\label{fig:flat}Left: Wave packet of the flat slicing, where $\sigma = C = 1$, $A = 0.1$, and $H = \sqrt{1/3}$. The blue contour is the classical trajectory. Right: Time slices of the flat slicing of the De Sitter space. Around the red-colored region, the wave function vanishes.}
\end{center}
\end{figure}

The on-shell solution is
\begin{eqnarray}
\varphi = \frac{\phi}{\sqrt{4\pi}} &=& \frac{A}{\sqrt{4\pi}}\frac{1}{a^{3}},\\
&=& - \frac{1}{4\pi \sqrt{3}} \sinh^{-1} \left( \sqrt{\frac{4\pi A^{2}}{3 H^{2}}} e^{- 12\pi \alpha/\sqrt{3}} \right),
\end{eqnarray}
where this is asymptotically ($\alpha \rightarrow -\infty$) satisfies $\varphi - \alpha = \mathrm{const.}$ contour. Therefore, this wave packet is consistent with the classical trajectories for the $\alpha < 0$ limit. The left of Fig.~\ref{fig:flat} shows that the wave packet is consistent with the classical trajectory for $\alpha < 0$, but as the wave reaches the potential well ($\alpha \rightarrow 0$), the coherence is lost due to the perfect absorption \cite{Kiefer:2004xyv}.

Now we can ask what is going on. The wave function is annihilated, but it is hard to believe that there is no future for the De Sitter space. In fact, we are living in some part of the $\alpha \gg 1$ limit (right of Fig.~\ref{fig:flat}). In this sense, we cannot say that there is no physical reality even if the wave function decays to zero.

On the other hand, we can ask what is happening in our expanding universe. What we can expect is that the universe should enter a state with accumulated quantum effects as it becomes larger and larger; the entire universe might be a superposition of diverse classical universes \cite{Hartle:2015bna}. What the wave function tells us is that the wave function is no longer a coherent state; rather, it should be a superposition of several classical geometries. We believe that the same thing should happen for the interior of the black hole. 

The common feature of the black hole and De Sitter spaces is that both of these systems are at their limit of information capacity (in the specific sense of achieving the maximum entropy allowed by the holographic principle for their size).
It appears that such systems must violate classical evolution (and possibly even quantum). Since such systems are usually maximally mixed, and in the black hole and De Sitter case the relevant degrees of freedom are gravitational, it is not surprising to get a superposition of several classical geometries.

\begin{figure}[t]
\begin{center}
\includegraphics[scale=0.45]{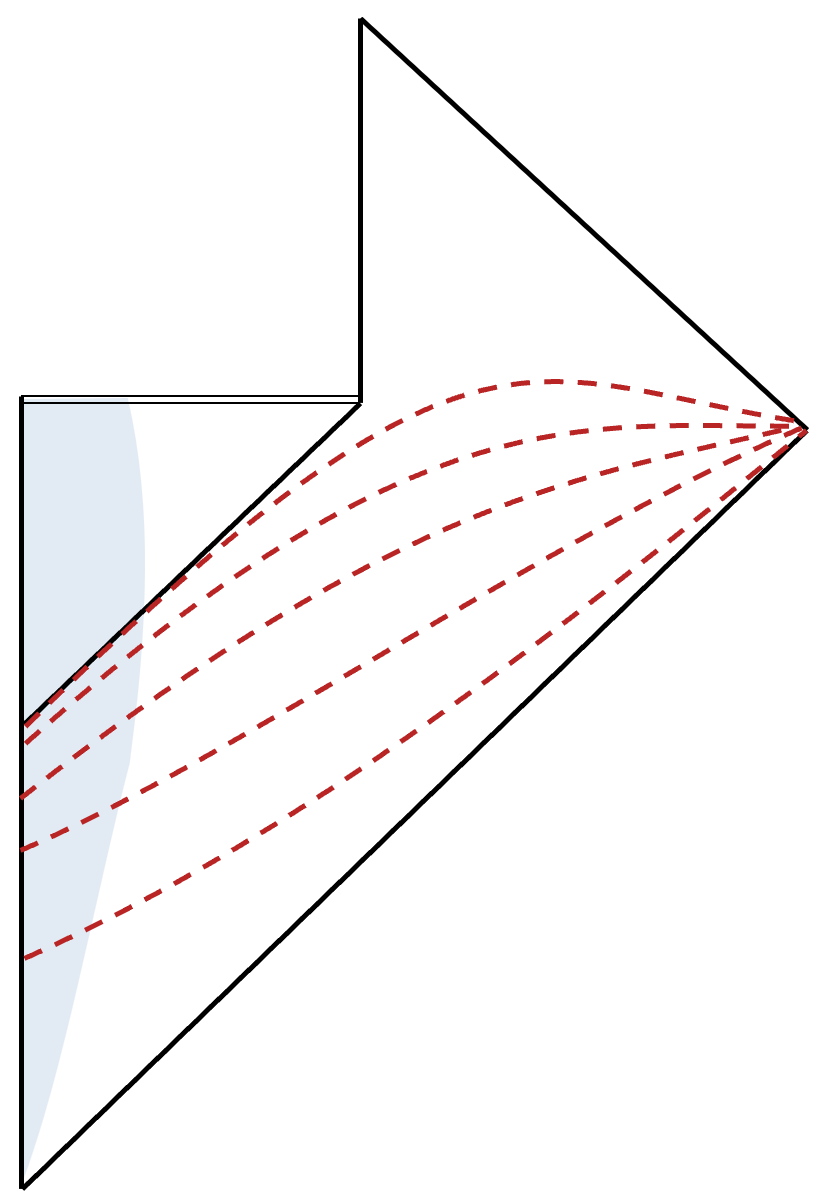}
\caption{\label{fig:allowed}Allowed time slices for an asymptotic observer keeping the wave function as coherent states.}
\end{center}
\end{figure}

\subsubsection{\textit{Nothing} as a superposition of coherent states}

Now, let us come back to the black hole case. We can think of several quantum states as follows.
\begin{itemize}
\item[--] \textit{Eigenstates}: In principle, one can classify the eigenstates of the Wheeler-DeWitt equation (after the separation of variables), but it is time independent, and hence, an eigenstate itself cannot present a classical spacetime.
\item[--] \textit{Coherent states}: This is a superposition of eigenstates for a classical spacetime. By imposing overlaps between different coherent states, one can explain their time evolution, which is consistent with the semi-classical evolution \cite{WSYMF}.
\item[--] \textit{Nothing}: By nothing, we mean the situation that the wave function vanishes \cite{Bouhmadi-Lopez:2019kkt}. However, as we observed in the De Sitter example, this does not mean that there is truly nothing. This is a quantum fluctuation, or a superposition of semi-classical states. In other words, it is a superposition of several coherent states.
\end{itemize}
Hence, one can interpret the annihilation-to-nothing results in a region that a single classical spacetime cannot describe. Instead, the spacetime must be a superposition of several classical spacetimes.

\subsection{Unitary time evolution of the asymptotic observer}

We can now construct the entire picture relevant to the information loss paradox and unitary time evolution. First, from our investigations, we know that we cannot choose a series of time slicings as we did in Fig.~\ref{fig:typicalslicing}. Instead, we should restrict ourselves to the time slices only outside the event horizon to have the coherent state at all times (Fig.~\ref{fig:allowed}).

\begin{figure}[h]
\begin{center}
\includegraphics[scale=0.75]{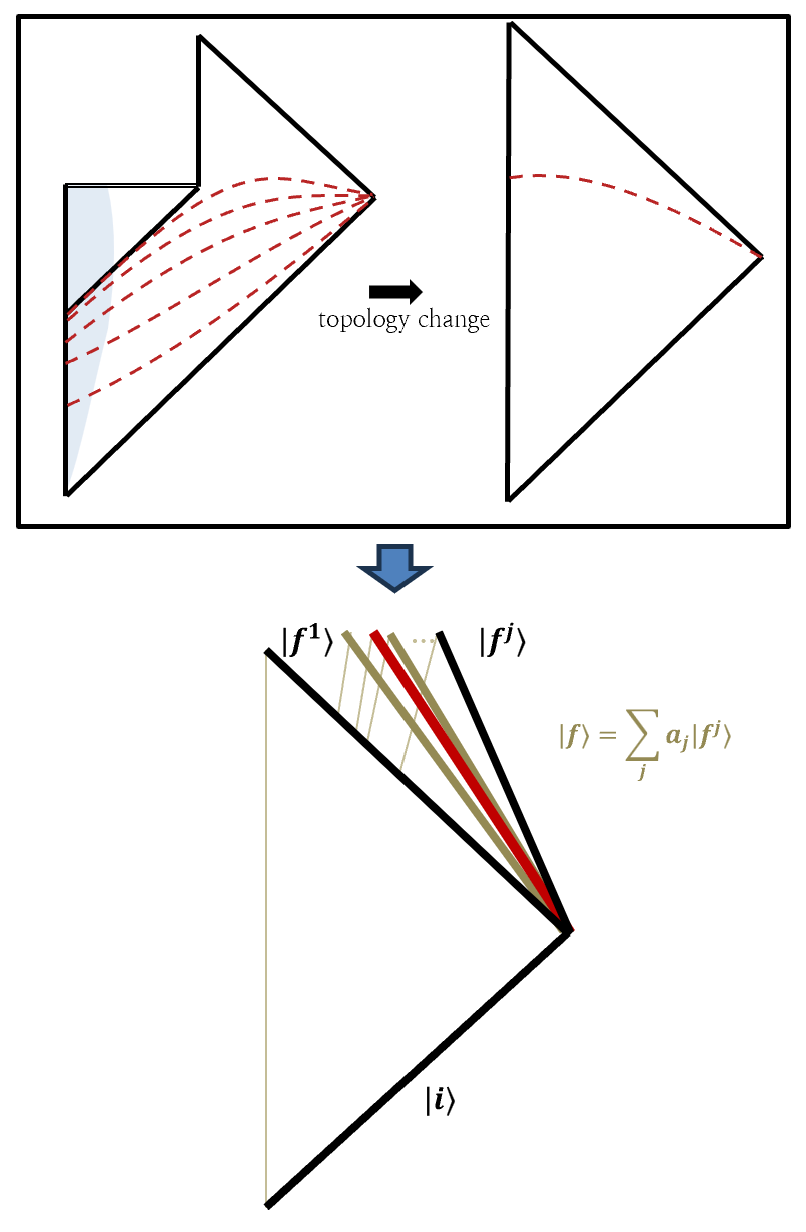}
\caption{\label{fig:topch}Conceptual picture of the topology change.}
\end{center}
\end{figure}

However, this is not sufficient to describe the unitary time evolution. What we can expect is that there may be a non-perturbative channel leading to a Minkowski geometry (Fig.~\ref{fig:topch}) \cite{Maldacena:2001kr,Hawking:2005kf}. We discuss some details as follows.
\begin{itemize}
\item[--] Thanks to the Euclidean path integral approach \cite{Hartle:1983ai}, one can construct an instanton connecting from a hypersurface of a semi-classical time slicing (Fig.~\ref{fig:allowed}, or equivalently, upper left of Fig.~\ref{fig:topch}) to that of Minkowski (upper right of Fig.~\ref{fig:topch}) \cite{Chen:2018aij}. Typically, the transition probability is $\propto e^{-S}$, where $S$ is the entropy of the black hole at some moment.
\item[--] Initially, the transition rate is exponentially suppressed, but as time goes on, $S$ decreases, and the transition probability increases. After careful computations, we can demonstrate that the transition to Minkowski is ultimately dominated \cite{Chen:2022ric}.
\end{itemize}

Therefore, as shown at the bottom of Fig.~\ref{fig:topch}, we have to sum over all possible non-perturbative transitions starting from an arbitrary initial time, and the resulting future boundaries might be entirely different ($| f^{j} \rangle$). Hence, the future boundary might be a superposition of several classical boundaries:
\begin{eqnarray}
| f \rangle = \sum_{j} a_{j} | f^{j} \rangle,
\end{eqnarray}
where $a_{j} = \langle f^{j} | i \rangle$ with a given in-state $| i \rangle$. This picture shares the same philosophy as \cite{Hartle:2015bna}.

\begin{figure}[h]
\begin{center}
\includegraphics[scale=0.75]{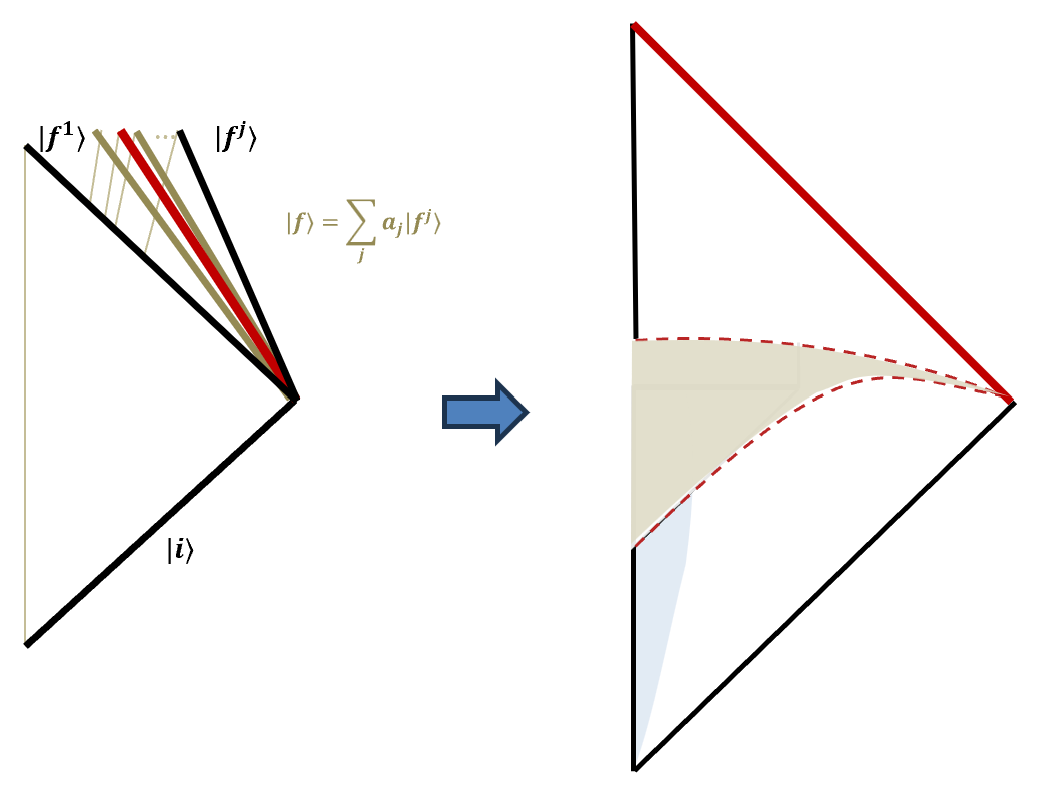}
\caption{\label{fig:unitarypatch}After the topology change, we can draw the most dominant history.}
\end{center}
\end{figure}

However, from computations in \cite{Chen:2022ric}, we can see that there exists a most dominant future boundary (red-colored line at the bottom of Fig.~\ref{fig:topch}). If we consider the most dominant history, it is reasonable to draw a picture as in Fig.~\ref{fig:unitarypatch}. This history indicates that the topology change occurs at a very late time; the time slicings before this change correspond to those of the semi-classical black hole, covering only the region outside the event horizon. 

As a result, there is, effectively, no event horizon. The time slice before and after the transition (gray-colored region in Fig.~\ref{fig:unitarypatch}) is formally connected by the Euclidean propagator:
\begin{eqnarray}
\langle f^{j} | i \rangle = \int_{i \rightarrow f^{j}} \mathcal{D} g_{\mu\nu} \mathcal{D} \phi \; e^{-S_{\mathrm{E}}}.
\end{eqnarray}
Therefore, it is not surprising that there is no information loss in the right of Fig.~\ref{fig:unitarypatch} (as was discussed in many scenarios which have no event horizon \cite{Saini:2015dea,Baccetti:2016lsb,Ayon-Beato:1999kuh,Ashtekar:2005cj,Haggard:2014rza}). We can choose all slices to be coherent states, making the picture self-consistent.

However, the remaining question is this: what happens if we choose an infalling observer? What is the physical reality inside the black hole?

\subsection{Infall problem: introducing an infall clock}

\begin{figure}[h]
\begin{center}
\includegraphics[scale=0.75]{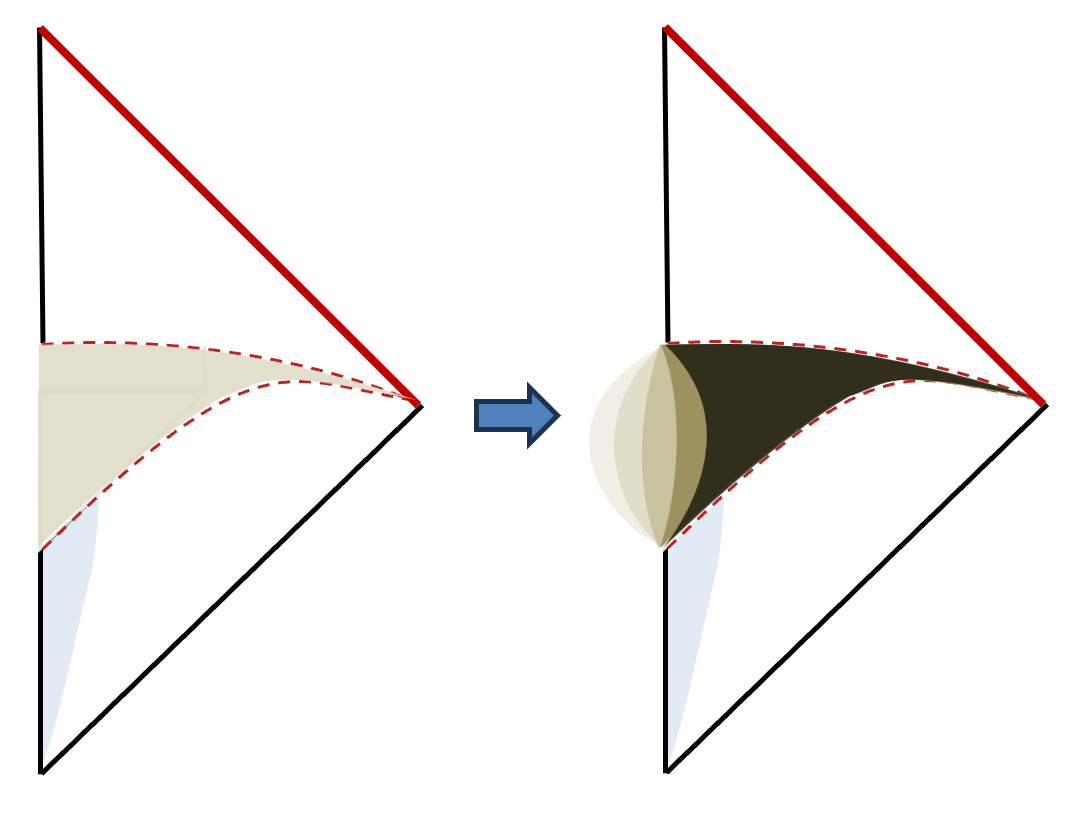}
\caption{\label{fig:center}As we see details of the interior of the black hole, a single coherent state cannot describe the interior. That means, a superposition of coherent states describes the interior of the black hole; this is the correct meaning of the vanishing wave function, i.e., annihilation-to-nothing.}
\end{center}
\end{figure}

As we computed, the interior of the black hole is nothing, i.e., the wave function should vanish. However, as we discussed, this \textit{nothing} is a superposition of coherent states (Fig.~\ref{fig:center}).

Then, one might further inquire about what is happening if there is an infalling observer. At first glimpse, thanks to the equivalence principle, there is no reason to deny that there can be an infalling observer who has semi-classical slicings. However, the question is, how can this observer define his own time? As we can see from the specific examples in \cite{Hajicek:1992pu,Vaz:2022jzq}, the choice of the time variables is essential to describe the infalling observer.

\subsubsection{Meaning of incommensurability between clocks}

One may ask what the meaning of changing clocks is. First, let us choose a clock at infinity and a corresponding series of time slices. When switching to a new time, one needs to introduce a new clock and trace out its clock degrees of freedom to apply the Page-Wootters formalism. This will result in a functional Schr{\"o}dinger equation with a new time variable. However, if it is not possible to obtain the functional Schr{\"o}dinger equation which is independent of the clock state, then the functional Schr{\"o}dinger equation itself cannot describe a unitary time evolution. In this case, we define that the times measured by the original and new clocks are \textit{incommensurable}.


Then, are the infinity and infalling clocks really incommensurable? We can explain this as follows, based on our computations.

\subsubsection{Incommensurability between infalling and asymptotic clocks}

Once we select such an infalling clock, it is then possible to consider a Gaussian wave packet for the interior of the black hole. However, as we have already shown, the interior of the event horizon cannot be described by a coherent state in terms of the clock at infinity. Therefore, the infalling observer's clock is a \textit{new physical clock}, which is incommensurable to the asymptotic clock.

As the infalling observer defines/constructs a new clock, this clock must be entangled with the whole system (of many possible geometries); hence, as one applies the Page-Wootters formalism, its quantum state already loses unitarity in terms of the asymptotic observer. Of course, this does not spoil the unitarity of the asymptotic observer, because the contribution of the infalling observer will be zero as the amplitude of the wave function of the asymptotic observer vanishes around the event horizon.


\subsubsection{What an infalling observer supposed to experience}

What will physically happen to the observer crossing the horizon? It has been reported that there appears to be an annihilation-to-nothing behavior inside the horizon \cite{Bouhmadi-Lopez:2019kkt}. For example, if we think about the Einstein gravity with a $U(1)$ gauge field model, one can choose a Gaussian wave packet defined at the event horizon that evolves to nothing around $r \sim M$ \cite{Chien:2025tzm}. Therefore, in this case, the infalling observer should change a clock once again around the $r \sim M$ surface. However, this does not complete the process; according to the BKL conjecture \cite{Belinsky:1970ew}, due to the chaotic nature near the singularity, an observer's local geometry might be well approximated by a Friedmann-Robertson-Walker metric. According to some computations, it is known that its wave function will vanish within a finite proper time \cite{Perry:2021mch}. Such a vanishing behavior near the singularity is consistent with several models with collapsing matter \cite{Saini:2014qpa,Greenwood:2008ht,Hajicek:2001yd,Brahma:2021xjy}.

This means that inside the black hole, there is a highly quantum regime. An infalling observer can reach the singularity within a finite proper time, according to general relativistic computations. However, quantum mechanically, the infalling observer should change their clock, and hence, the loss of unitarity is cascaded.

\subsection{Generalized Uncertainty Principle revisited}

Is there any way to constructively understand the black hole as nothing, i.e., as a superposition of various coherent states? Perhaps one can think of the fuzzball picture \cite{Mathur:2005zp}. If one can construct a vast number of inequivalent configurations, it will be helpful to understand the black hole in a unitary observer's sense. However, it is fair to say that the fuzzball picture strongly relies on the detailed matter contents of the string theory. Can there be any more straightforward explanation relying just on gravity?

In this regard, one can gain insight from the Generalized Uncertainty Principle (GUP) \cite{Hossenfelder:2012jw}. Several quantum gravitational models provide an idea that there exists a fundamental minimal length scale, so to speak, $\ell_{P}$. Due to this, apart from the typical uncertainty principle of quantum mechanics, the uncertainty of length scales might increase as we increase the probing energy scale, where such a bouncing moment is the length scale at $\ell_{P}$. Mathematically, one can write as follows:
\begin{eqnarray}
\Delta x \Delta p \geq \frac{1}{2} \left( 1 + \alpha  \ell_{P}^{2} \left( \Delta p \right)^{2} \right).
\end{eqnarray}
By adapting this, we can conclude that there exists a minimal length scale, in the sense that one cannot decrease the uncertainty below that value, even if we increase the probing energy. However, in any case, the physical application of such a minimal length will be applicable only if the length scale is of the order of $\ell_{P}$.

However, apart from this common sense of GUP, mathematically, the relation states that the uncertainty will monotonically increase as the probing energy $\Delta p \simeq M$ increases. Therefore, if $\Delta p \gg 1$,
\begin{eqnarray}
\Delta x \propto \Delta p \simeq M.
\end{eqnarray}
Hence, the GUP literally states that the black hole can have an order $M$ uncertainty, or \textit{the horizon scale uncertainty}, if we interpret the gravitational collapse as a probing process.

It is interesting that both the GUP and our picture point toward the existence of a horizon scale uncertainty; in other words, a superposition of semi-classical states exists at the horizon scale. Is this an accident, or is there a more fundamental relationship? This question goes beyond the scope of the present draft, but there is a reasonable guess. Perhaps, the horizon scale annihilation-to-nothing behavior is a macroscopic phenomenon, but its origin might be microscopic; a condensation of microscopic effects can change macroscopic behaviors, as we have observed in many examples of quantum mechanics. Quantum mechanics is a microscopic theory, but there exist many macroscopic phenomena that cannot be explained without assuming the principles of quantum mechanics (e.g., superconductivity and superfluidity).

\subsection{Comparison to the firewall controversy}

Finally, we remark on the consistency between different observers. In terms of the unitary observer, the whole evolution is unitary and consistent with the semi-classical physics, but there exists a horizon scale uncertainty; this means that we cannot choose a time slice that crosses over the horizon. On the other hand, if there is an infalling observer, one needs to select an infalling clock, which already breaks the unitarity. Hence, among the five assumptions mentioned in the introduction, a single observer cannot keep everything. The unitary observer lost general relativity of the horizon scale, while the infalling observer lost the unitarity of quantum mechanics. This helps us avoid inconsistencies in the black hole complementarity \cite{Susskind:1993if} and the firewall controversy \cite{Almheiri:2012rt} (see also \cite{Hutchinson:2013kka}).


The final question we address here is this: By staying outside the horizon for all time, are we clearly unitary observers? This is a subtle question, but the best we can say is that we are mostly unitary observer. As we see at the bottom of Fig.~\ref{fig:topch}, the Euclidean path integral indicates that the final boundary of the unitary observer must be a superposition of several classical boundaries. However, almost all of them are not dominant. Only if the subdominant contributions are negligible, can we say that we are an outside observer at the right of Fig.~\ref{fig:unitarypatch}.

Therefore, we may not be perfect unitary observers, and hence we probably lose some small amount of information effectively in terms of Fig.~\ref{fig:topch} \cite{Hawking:2014tga}. However, we can also recover some of it, as shown in Fig.~\ref{fig:unitarypatch}. This finding is actually consistent with quantum mechanics, where definite classical statements very often do not make much sense. 

\section{\label{sec:dis}Discussion}

In this paper, we discussed the information loss paradox in the light of canonical quantum gravity. The wave function of the universe carries all the information but does not have a global time, as in Wheeler-DeWitt equation. To demonstrate the unitary time evolution, we introduced the time following the Page-Wootters formalism, i.e., by tracing out the degrees of freedom of a clock. The unitary time evolution does make sense as long as the clock is located at infinity.

However, introducing time is not sufficient. We also need to introduce a series of spacetime hypersurfaces. Then, the question is whether a hypersurface can cross the event horizon. We cannot answer this question  unless we solve the wave equation (otherwise, we are prejudicing an answer), but to define the wave equation, we need to determine the time slice. Even with this logical loop, we can still investigate the self-consistency condition by requiring that we should be able to assign a coherent state for the hypersurface, i.e. a Gaussian wave packet along the classical contour.

But, can we provide a Gaussian wave packet for a time slice that crosses over the event horizon? To answer this, we wrote the Wheeler-DeWitt in terms of the metric ansatz that crosses over the event horizon. Although this equation is a functional differential equation, it was still possible to trace the existence of coherent states. We concluded that we could provide a coherent state only outside the event horizon, while inside the wave function had to vanish. This can be viewed as a realization of the annihilation-to-nothing at the horizon scale.

However, we have to interpret what this vanishing of the wave function (or annihilation to nothing) really means. As we demonstrated in a De Sitter example, the vanishing wave function does not imply that there is really nothing left; instead, there is something that is a superposition of several coherent states, which a single coherent state cannot describe. This fact is closely related to the horizon scale uncertainty, as indicated by the generalized uncertainty principle.

Therefore, a unitary observer can indeed exists, as long as all the time slices are outside the event horizon, and the geometry eventually transits to a topologically trivial geometry (through the Euclidean propagator). In this sense, effectively the event horizon is never encountered, and hence, no loss of information. On the other hand, an infalling observer will experience classical dynamics according to general relativity, but this observer needs to introduce an infalling clock, which implies that the observer evolves into a mixed state (as seen by an asymptotic observer). In terms of the unitary observer, the contribution of the infalling observer will be negligible, as the wave function inside the horizon vanishes.

While this paper offers a new perspective on understanding the information loss paradox, we suggest several questions that may shed more light on this point of view as follows.
\begin{itemize}
\item[--] What is the dynamical mechanism to increase the uncertainty inside the black hole? Can microscopic physics be connected to this macroscopic phenomenon?
\item[--] Can there be any experimental model (e.g. analog gravity) that demonstrates this kind of behavior?
\item[--] Is this annihilation-to-nothing behavior universal? Is there any dependence on models, e.g., various dimensions or matter fields? What about the cosmological horizons? 
\end{itemize}
We leave these interesting questions for future investigations.

\section*{Acknowledgment}
WL was supported by the National Research Foundation of Korea, Basic Science Research Program under the Grant No. RS-2024-00336507.
DY was supported by the National Research Foundation of Korea (Grant No.: 2021R1C1C1008622, 2021R1A4A5031460).
DS was partially supported by the US National Science Foundation under the
Grant No. PHY-2310363.

\end{document}